\documentstyle[aps,prb,epsfig,twocolumn]{revtex}
\begin{document}
\draft
\title{LDA+DMFT study for $\rm La_{1-x}Sr_xTiO_3$}
\author{M.\ B.\ Z\"olfl, Th.\ Pruschke, J.\ Keller}
\address{Institut f\"ur Theoretische Physik I, Universit\"at
Regensburg, Universit\"atsstr. 31, 93053 Regensburg, Germany}
\author{A.\ I.\ Poteryaev, I.\ A.\ Nekrasov, V.\ I.\ Anisimov}  
\address{Institute for Metal Physics, 620014 Ekaterinburg, Russia}
\date{\today}
\maketitle

\begin{abstract}
The dynamical mean-field theory together with
the non-crossing approximation is used to set up a novel scheme to study
the electronic structure of strongly correlated electron systems. 
The non-interacting band structure is obtained from a density
functional calculation within the local density approximation.
With this method the doped Mott insulator $\rm La_{1-x}Sr_x Ti O_3$ is
studied. Starting from first-principle calculations for a cubic and an
orthorhombic system we determined the one-particle spectrum.
Both one-particle spectra show a lower Hubbard band 
(seen as $d^1 \rightarrow d^o$ transitions in photo emission experiments),
a Kondo resonance near the Fermi energy and the upper Hubbard
band ($d^1\rightarrow d^2$ transitions in an inverse photoemission experiment).
The upper Hubbard band develops a multipeak structure, a consequence of the 
consideration of all local two-particle correlations, which leads to the 
full multiplet structure in the atomic limit.
The calculation for the orthorhombic system shows qualitative good agreement 
when compared with experimental photoemission spectra.
\end{abstract}

\section{Introduction}               
\label{intro}
One of the most challenging problems in condensed-matter physics is the 
accurate calculation of the electronic structure starting from first 
principles, since it is impossible to solve the many-body problem without
severe approximations. A great success in first principle methods 
was the development of the density functional theory (DFT)\cite{LSDA}, 
which produces an exact groundstate energy and maps the many-body problem onto 
a ficticious single-electron problem with an one-electron exchange 
and correlation potential.
The method would be exact, if one knew this particular potential.

A commonly used approximation replaces this functional by the one
of the homogeneous electron gas. This approximation is widely known 
as the local (spin-) density approximation (L(S)DA) \cite{LSDA}. 
The LDA turns out to be very crude, if the on-site Coulomb interaction 
between electrons becomes very strong, i.e. if the interaction strength 
is of the order of, or larger than the bandwidth. 
This is particular the case in systems with
$d$- or $f$-electrons. As is well known, the L(S)DA approximation leads to 
wrong results for stronlgy correlated electron systems such as
Mott-Hubbard insulators or heavy fermion systems.

A first step towards a better description of strongly correlated
systems was the introduction of the LDA+U method \cite{LDA+U}, a
Hartee-Fock-like scheme, that is able to account for a variety of
interesting effects in transition metal compounds. For example, the
LDA+U allows to describe correctly the groundstates of magnetically, 
charge and orbitally ordered materials \cite{LDA+Usuccess}. 
Due to its Hartree-Fock nature, however, this approximation still has 
severe drawbacks.

To go beyond LDA+U, a new scheme was recently proposed \cite{AniKot}, 
which combines the LDA bandstructure calculation with the dynamical 
mean-field theory (DMFT: for a review see reference 5). The DMFT maps
the lattice model onto an effective impurity problem and
can be used to solve the many-body problem self-consistently with
various methods, like e.g.\ quantum Monte Carlo (QMC) \cite{QMCDMFT}, 
exact diagonalization (ED) \cite{EDDMFT}, 
iterated perturbation theory (IPT) \cite{IPTDMFT} or within the 
non-crossing approximation (NCA) \cite{NCADMFT}.

A calculation scheme for LDA+DMFT in conjection with IPT was already accomplished
in reference 4. However, in this work only one
mean Coulomb parameter $U$ was used to describe all local correlations.
It is however well known that the complexity of the physics observed
in e.g.\ transition metal oxides is governed by the subtle interplay
of the various local Coulomb integrals, i.e.\ the neglect of the
detailed orbital structure presents a serious limitation of the
method.

In this paper a LDA+DMFT approach with the NCA is introduced,
which allows to consider the full local multiplet structure. Thus, an
intra orbital Coulomb energy $U$, an inter-orbital Coulomb energy
$U^\prime$, and Hund's coupling constant J are considered within a
SU(N) invariant formulation of the local Hamiltonian for degenerated
bands \cite{MOHM}. The bandstructure calculations as starting point
for the DMFT were done with the linearized muffin-tin orbital method
(LMTO) \cite{LMTO}.
The paper is organised as follows. In section \ref{scheme} we
introduce the Hamiltonian used for the many-body calculation and describe 
the details of the NCA scheme used to solve the local problem .
Results for $\rm La_{1-x}Sr_xTiO_3$ as a particularly interesting
example of a correlated metal obtained with this technique are
collected in section \ref{calc} and a summary in section \ref{summary} 
will conclude the paper.
   
\section{The calculating scheme}
\label{scheme}
In order to combine the achievements of the DMFT with the framework of
DFT, one needs a calculation sche\-me that allows to
map the calculated band structure onto a suitable tight-binding
model. This can be done most conveniently with the LMTO in the orthogonal
approximation \cite{LMTO}. With this technique a tight-binding model can 
naturally be constructed from the DFT+LDA bands in real-space representation
\begin{eqnarray}
H_{tb}=\sum_{ilm,jl^\prime m^\prime,\sigma} &&
(\delta_{ilm,jl^\prime m^\prime}\epsilon_{il}\hat{n}_{ilm\sigma}\nonumber\\
&&+ \quad t_{ilm,jl^\prime m^\prime}
\hat{c}^\dagger_{ilm\sigma}\hat{c}_{jl^\prime m^\prime \sigma})\;\;.
\end{eqnarray}  
Since the LDA one-electron potential is orbital independent and 
takes into account the Coulomb interaction in an averaged way, 
one can easily generalize this Hamiltonian by adding local Coulomb 
correlations:
\begin{eqnarray}
&&H_{corr}=\nonumber\\
&&\phantom{+}\:\frac{1}{2}
\sum_{il,m\sigma
m^\prime\sigma^\prime}\!\!\!\!\!\!\!\!\!\!\!\!\!\phantom{\sum_{m}}^{\prime}\:
U^{il}_{mm^\prime}\:\: \hat{n}_{ilm\sigma}\hat{n}_{ilm^\prime\sigma^\prime}\\
&&+\:\frac{1}{2}\sum_{il,m\sigma m^\prime\sigma^\prime n\rho
n^\prime\rho^\prime}\!\!\!\!\!\!\!\!\!\!\!\!\!\!\!\!\!\!\!\phantom{\sum_{m}}^{\prime}
\quad
J^{il}_{m\sigma m^\prime\sigma^\prime n \rho n^\prime \rho^\prime}\:\: 
\hat{c}_{ilm\sigma}^\dagger\hat{c}_{ilm^\prime\sigma^\prime}^\dagger
\hat{c}_{iln\rho}          \hat{c}_{iln^\prime\rho^\prime}.
\end{eqnarray}
In term (2) and (3), the index $i$ labels the unit cell, $l$ denotes
the atom in the unit cell and $m\sigma$ and others stand for orbital and spin.
The prime on the sum indicates that at least two of the indices on different
operators have to be different to account for Pauli's principle.
In our approach we are able to include all local two-parcticle 
correlations appearing in term (2) and (3). 
In the following we will assume, that it is only necessary to take
into account the Coulomb interactions for the $d$-shell of the 
transition metal ions ($i=i_d$ and $l=l_d$) explicitly.
Therefore the indices $il$ will be omitted.
All other orbitals will be considered as itinerant bands, which are well
described by the LDA.

Let us discuss the interaction term in more detail. 
The first term (2) describes the density-density Coulomb 
repsulsion. Here, we introduce two distinct Cou\-lomb parameters:
the intra-orbital Coulomb energy $U$ has to be considered in case of a doubly 
occupied orbital, while the inter-orbital Coulomb energy $U^\prime$
applies for example in the case of a doubly occupied $d$-shell
with electrons on $d$-orbitals with different indices.
The second term (3) describes Hund's coupling and all other 
off-diagonal interactions. 
Although in principle the full Coulomb matrix (2) and (3) can be used, 
we restrict ourselves for practical purposes to the
following, commonly used form for the two-particle interactions of the
$d$-electrons
\begin{eqnarray}
&&H_{corr}=\nonumber\\
&&\phantom{-}U \sum_m \hat{n}_{m\uparrow}\hat{n}_{m\downarrow}+
\frac{U^\prime}{2}  \sum_{m,m^\prime,\sigma,\sigma^\prime}^{m\ne m^\prime} 
              \hat{n}_{m\sigma} \hat{n}_{m^\prime\sigma^\prime}\\
&&-\frac{J}{2}  \sum_{m,m^\prime,\sigma}^{m\ne m^\prime} 
              \hat{n}_{m\sigma} \hat{n}_{m^\prime\sigma}
+ J \sum_{m,m^\prime}^{m\ne m^\prime} 
    c^\dagger_{m\uparrow}c^\dagger_{m^\prime\downarrow}
    c_{m\downarrow}         c_{m^\prime\uparrow}\\
&&+ J_C \sum_{m,m^\prime}^{m\ne m^\prime} 
    c^\dagger_{m\uparrow}c^\dagger_{m\downarrow}
    c_{m^\prime\downarrow}         c_{m^\prime\uparrow},
\end{eqnarray}
with a density-like Coulomb term (4), a SU(2)-invariant form of Hund's 
Coupling (5) and a charge-flip term or pair-\-hopping term (6). 
$J_C=J$ is chosen here as a good approximation. 

Since the LDA already contains the influence of the Coulomb
interaction to a certain degree, the problem of double counting of
these contributions by using $H_{corr}$ arises. In order to avoid this 
double counting one has to subtract off the interaction contributions from the
LDA. Unfortunately, the precise form for one particular set of
orbitals is not known and the best one can do is to account for these
contributions in an averaged way. Since the LDA total energy is to a good 
approximation a function of the total number of electrons, one can assume 
that the interaction contributions have the form
\begin{equation}
E_{I}=\frac{1}{2} \bar{U} n_d (n_d-1) 
- \frac{1}{2}J\sum_\sigma n_{d,\sigma} (n_{d,\sigma}-1)\;\;.
\end{equation}
Here, $\bar{U}$ is the mean value for the Coulomb interaction and may be
obtained from a first-principle calculation \cite{Parameters} or
from experiment, for example high-energy spectroscopy.
$n_d$ is the total number of $d$-electrons and $n_{d,\sigma}$ is the number 
of $d$-electrons with spin $\sigma$. 
Distinguishing intra- and inter-orbital interaction for a 
$N_{deg}$-fold degenerate orbital-system one can determine $U$ and $U^\prime$ 
by the relation
\begin{equation}
\bar{U}=\frac{U + 2(N_{deg}-1) U^\prime }{(2N_{deg}-1) }\;\;.
\end{equation}    
Starting from given values for $\bar{U}$ and J and using the relation
$U=U^\prime+2J$, which is a consequence of the rotational invariance
of the Hamiltonian \cite{MOHM}, one is able to determine the
intra- and inter-orbital Coulomb interaction $U$ and $U^\prime$.
Finally, only an expression for the local energy of the $d$-states is needed.
In the spirit of L(S)DA+U we can define such a one-electron energy by
\begin{eqnarray}
\varepsilon^0_{d}&\quad&=\: \frac{d}{dn_{d \sigma}}(E_{LDA}-E_{I})\\
&\quad&=\: \varepsilon^{LDA}_{d}
  -\bar{U}(n_d-\frac{1}{2})
  +\frac{J}{2}(n_d-1)\nonumber\\
&\quad&=\: \varepsilon^{LDA}_{d}
 -\frac{U+2(N_{deg}-1)U^\prime}{2N_{deg}-1}(n_d-\frac{1}{2})
  \nonumber\\
&\quad&  \displaystyle+\frac{J}{2}(n_d-1),
\end{eqnarray}
with
\begin{equation}
\varepsilon^{LDA}_{d}=\frac{d}{dn_{d \sigma}}E_{LDA}
\end{equation} 
and $E_{LDA}$ the total energy as calculated from DFT+LDA. With the corrected 
total energy functional $E_{LDA}-E_{I}$ one has to construct the 
tight-binding Hamiltonian within the framework of the LMTO-method \cite{LMTO}.

Given the band structure we are now in the position to set up the scheme 
necessary for the DMFT. First, let us define the non-interacting Green 
function
\begin{equation}
G^0_{\sigma}(z)=\int d\omega \frac{\rho^0_{\sigma}(\omega)}{z-\omega}
\end{equation}
via the Hilbert transform of the spectral function $\rho_{\sigma}^0(\omega)$
obtained from the Hamiltonian (1). Note that in general $\rho_{\sigma}^0(\omega)$
and consequently $G^0_{\sigma}(z)$ will be matrices in orbital space. The most
important feature of the DMFT is that the proper one-particle self energy due
to the local Coulomb interaction is purely local \cite{DMFT}. Thus, we obtain as an
expression for the full Green function of the interacting system
\begin{equation}
G_{\sigma}(z)=G^0_{\sigma}(z-\Sigma(z))
=\int d\omega \frac{\rho^0_{d \sigma}(\omega)}{z-\Sigma_{\sigma}(z)-\omega}\;\;.
\end{equation}
This equation must again be read as a matrix relation in orbital space.
In the spirit of the DMFT \cite{DMFT} eq.\ (13) can be cast into the form
\begin{equation}
G_{\sigma}(z)=
\frac{1}{z-\varepsilon^0_{d}-\Sigma_{\sigma}(z)-\Delta_{\sigma}(z)}
\end{equation}
where a so-called hybridization function was introduced, which fulfills the
standard relation
\begin{equation}
\lim_{\omega\rightarrow\pm\infty}\Re e\{\Delta_{\sigma}(\omega+i\delta)\}
=0\;\;.
\end{equation}
The one-electron energy $\varepsilon^0_{d}$, the imaginary part of the 
hybridization function and the local Coulomb parameters define a multi-orbital
impurity model, which has to be solved to obtain the selfenergy 
$\Sigma_{\sigma}(z)$.
Finally, eqs.\ (13), (14) and the solution of this impurity model define a 
self-consistency cycle for the solution of the lattice model within the DMFT.

In the following we will set up a resolvent perturbation theory for the
multi-orbital impurity model defined above. 
Within this approach the model is solved approximately by considering the 
lowest order diagrams only, which is the well-known NCA \cite{NCADMFT}. 
To this end we first have to diagonalize the local Hamiltonian
\begin{equation}
H_{loc}\quad = \quad \sum_{\alpha} E_\alpha |\alpha\rangle\langle\alpha|\;\;
\end{equation} 
and use the basis states $\left\{|\alpha\rangle\right\}$ in the further
procedure. As a consequence, all local interactions are fully included
in these impurity states. Therefore, all types of local interaction terms can
in principle be considered leading to a realistic multiplet structure in the
atomic limit.    
To account for the hybridization term, we
have to express the fermionic creation and annihilation operators for
the $d$-orbitals in terms of our basis  $\left\{|\alpha>\right\}$, i.e.
\begin{eqnarray}
  d^\dagger_{\kappa\sigma} &&=\sum\limits_{\alpha,\beta}
  D^{\kappa\sigma*}_{\beta\alpha} |\alpha\rangle \langle \beta|\:,\\
  d_{\kappa\sigma} &&=\sum\limits_{\alpha,\beta} D^{\kappa\sigma}_{\alpha\beta}
  |\alpha\rangle \langle \beta|\:,
\end{eqnarray}
with $\kappa$ the orbital index and $\sigma$ for the spin.
With (17) and (18) one can straightforwardly formulate the NCA equations
\cite{NCADMFT} as
\begin{equation}
R_{\alpha}(z)=R^0_{\alpha}(z)
+R^o_{\alpha}(z)\Sigma_{\alpha}(z)R_{\alpha}(z),
\end{equation} 
with
\begin{eqnarray}
 && R^0_{\alpha}(z)=\frac{1}{z-E_\alpha},\\
        &\quad\nonumber\\
 && \Sigma_{\alpha}(z)=-\frac{1}{\pi}\sum\limits_\sigma\sum\limits_{\kappa\kappa'}\sum\limits_{\alpha'}\int 
  d\varepsilon\nonumber\\
&&  \Im m\{\Delta^{\kappa\kappa'}_\sigma(\varepsilon)\}  
  {\rm D}^{\kappa\sigma*}_{\alpha,\alpha'}{\rm D}^{\kappa'\sigma}_{\alpha,\alpha'} f(\eta_{\alpha\alpha'}\varepsilon)
  R_{\alpha'}(z+\eta_{\alpha\alpha'}\varepsilon).
\end{eqnarray}
$f(\epsilon)$ is the Fermi function and
$\eta_{\alpha\alpha'}$ is equal to 1 or -1 depending on
wether one has to add or subtract an electron to excite the impurity from 
state $|\alpha\rangle$ to state $|\alpha'\rangle$. Again, eqs.\ (19)-(21) have
to be solved self-consistently.
Finally, we are able calculate the $d$-Green function from
\begin{eqnarray}
&&G^{\kappa\kappa'}_{\sigma}(i\omega) = \frac{1}{Z}\sum\limits_{\alpha,\alpha'}
D^{\kappa\sigma*}_{\alpha\alpha'}D^{\kappa'\sigma}_{\alpha\alpha'}
\oint dz \nonumber\\ 
&&\quad\quad\quad\quad\quad\quad\quad\quad
\frac{e^{-\beta z}}{2\pi i} R_{\alpha}(z)R_{\alpha'}(z+i\omega)\;\;.
\end{eqnarray}
In eq.\ (22), $\displaystyle Z=\sum\limits_\alpha\oint\frac{dze^{-\beta z}}{2\pi i}
R_\alpha(z)$ denotes the impurity partition function and $\beta$ is the inverse temperature.

The solution eq.\ (22) of the impurity problem closes the cycle (13) and (14):
Starting from an initial guess for the 
selfenergy $\Sigma_{\sigma}(z)$ one obtains from eq.\ (14)
the hybridization function $\Delta_{\sigma}(z)$, which is inserted into 
the NCA (21).
From the impurity Green function (22) and the hybridization function one can 
then calculate a new self energy from eq.\ (14) and repeats the procedure
until self consistency is reached.

\section{Results for $\rm La_{1-x}Sr_xTiO_3$}
\label{calc}
We applied the procedure discussed in the previous section to $\rm La_{1-x}Sr_xTiO_3$.
This
material is at high temperatures a paramagnetic metal for the undoped case 
(x=0) and becomes an antiferromagnetic insulator below $T_N=125K$ with a 
very small gap of $0.2 eV$. Doping with a few percent of $\rm Sr$ leads to 
a paramagnetic metal with a large effective mass \cite{Fujimori}. 
As photoemission experiments show, $\rm La_{1-x}Sr_xTiO_3$ may be regarded as
an example of a strongly correlated metal, i.e.\ the density of states is that
of a doped Mott-Hubbard insulator.

The $\rm LaTiO_3$ crystal is a slightly distorted cubic perovskite. The octahedral coordination of the oxygen ions leads to a $t_{2g}$-$e_g$ crystal-field 
splitting, such that both bands $t_{2g}$ and $e_g$ are well separated, 
as can be seen in a standard LDA calculation (LMTO method).
At the Fermi level one finds the $t_{2g}$ band of the $\rm Ti$ $3d$ states. $3$eV above the Fermi level and well separated from the $t_{2g}$ band the $e_g$ 
band is located.
Lead by these results we can restrict ourselves to the $3$-fold 
degenerate $t_{2g}$ orbitals as relevant for the low temperature properties. 

The Coulomb parameters $\bar{U}=4 eV,J=0.6 eV$ we use for the DMFT 
calculation were obtained from a constraint LDA calculation \cite{Parameters}.
$\bar{U}$ was determined in a calculation, where all electrons 
except $t_{2g}$ ones can screen the 
Coulomb interaction. (
The influence of the $e_g$ electrons on the Coulomb interaction
inside $t_{2g}$ subshell was investigated in \cite{solovyev,pickett}).
  In order to show the effect of varying the Coulomb
parameter on the results we chose $\bar{U}=6 eV$ in the first parameter 
sets, which refers to the case where $e_g$-electrons do not
participate in the screening.

\begin{figure}
{\epsfig{file=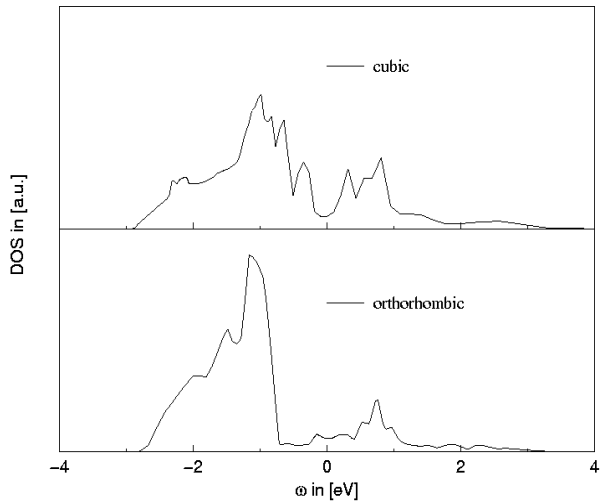,width=8cm}}
\vspace{0.5cm}  
\caption{The non-interacting $t_{2g}$ density of states for LaTiO$_3$ 
corrected according to eq.\ (7) with $\bar{U}=6 eV,J=0.6 eV$.
The upper part shows the density for a cubic, the lower part for an 
orthorhombic system.}    
\label{fig:1}
\end{figure} 
In figure \ref{fig:1} we show the non-interacting density of states for the 
$t_{2g}$ bands only taken from a LDA calculation for a cubic 
(upper part of figure \ref{fig:1}) and a orthorhombic system 
(lower part of \ref{fig:1}).       
Both spectra are corrected according to eq.\ (7).
From the DOS in figure \ref{fig:1} we can determine the one-particle energy 
$\epsilon^0_{t_{2g}}$ as average energy, and
the hybridization function $\Delta_{t_2g}(\omega)$ is then defined by 
equation (13) and (14). In figure \ref{fig:1} the  
bandwidth determined by the main contributions of the spectrum is 
varying for different underlying structures examined by the LDA 
calculation. The orthorhombic system leads to a narrower spectrum 
with the main contribution below the Fermi energy. The spectrum for the 
cubic system is broader and the spectral weight is spread over the 
whole range of the band. As a consequence, the average energy of the 
orthorhombic system is lower than for the cubic, which leads to different
one-particle energies in the resulting impurity model. 
It is evident that this change of bandwidth and of one-particle energy 
will eventually lead to different physics for both systems.
   
A particularly important simplification arises from the fact that the 
tight-binding Hamiltonian (1) constructed from the corrected LDA $t_{2g}$ bands
turns out to be diagonal in orbital space. This means that all orbital
matrices in eqs.\ (13), (14), (21) and (22) are diagonal and -- since we
do not allow for orbital ordering -- degenerate with respect to the orbital 
indices, which results in a tremendous reduction of the necessary 
computational effort.

\begin{figure}
{\epsfig{file=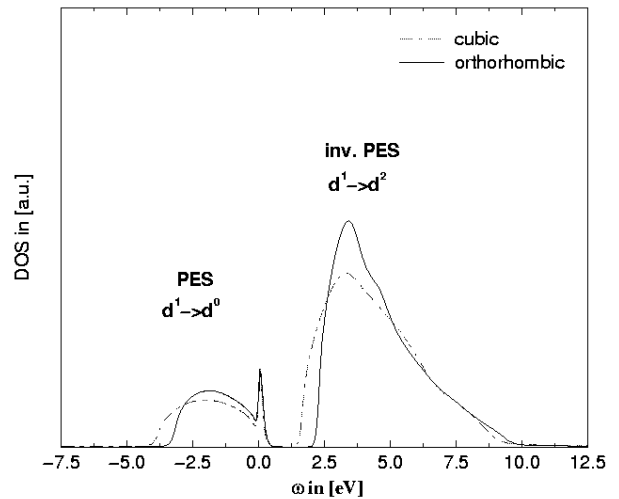,width=8cm}} 
\vspace{0.5cm} 
\caption{$t_{2g}$ density of states for La$_{1-x}$Sr$_x$TiO$_3$ for cubic and 
orthorhombic crystal structure. The parameters are:  $\bar{U}=6 eV$ and $J=0.6 eV$, a reciprocal temperature $\beta=100 \:eV^{-1}$ and a doping of $x=6$\%.
The denoted transitions are the removal of electron from a single occupied 
$t_{2g}$-state and the addition of an electron to a doubly occupied state.}    
\label{fig:2}
\end{figure} 
The results of the self-consistent DMFT calculation for the $t_{2g}$ spectral
function of a cubic and a orthorombic system with a partially screened 
mean Coulomb interaction of $\bar{U}=6 eV$ and $J=0.6 eV$ and a reciprocal temperature $\beta=100\:\:eV^{-1}$ are shown in figure \ref{fig:2}. 
The doping level was set to $x=6\%$. Both systems
lead to qualitatively similar spectra. Especially 
the lower Hubbard band reaches from low energies up to the Fermi 
level. This band refers to $d^1\rightarrow d^0$ transitions in a 
photoemission experiment. 
At the Fermi energy the well known many-body resonance can be seen, which
describes singlet-triplet excitations of the impurity and the surrounding 
effective medium \cite{Kondo}.  
At higher energies the upper Hubbard band occurs which refers to
$d^1 \rightarrow d^2$ transitions in an inverse photoemission 
experiment. Here contributions, which stem from transitions into several multiplet states, are almost completely smeared out to a broad, featureless band.
Apart from these gross features, the spectra show clear differences between
orthorombic and cubic system. Especially the influence of different bandwidth
is clearly visible, the narrower non-interacting spectrum for the orthorhombic
system leading to narrower peaks in the DMFT spectrum. As a consequence,
the value of the gap between lower and upper Hubbard band is increased, too.
Let us stress that the value of the gap, due to the complex atomic multiplet 
structure, is not simply given by the pure Coulomb interaction $\bar{U}$.
For a partially screened mean Coulomb value of $\bar{U}=6eV$ the real Coulomb 
parameters are $U=6.96 eV$ and $U^\prime=5.76 eV$.
Thus the lowest lying excitation is into a spin $S=1$ state with an excitation energy of $\epsilon^0_{t_{2g}}-(U^\prime-J)$. 
This would lead to a gap value of $5.16eV$ minus bandwidth.
The smaller bandwidth of the orthorhombic sytem thus turns out to 
produce an overall larger gap.     

In order to exploit the difference of the value, and hence the impact
of the procedure to determine $\bar{U}$, we compare the previous
results to a calculation where a fully screened mean Coulomb energy of 
$\bar{U}=4eV$ is used (i.e. $U=4.96 eV$ and $U^\prime=3.76 eV$).
Note that the value of $J$ is not changed by using different screening 
scenarios, which reflects the well-known fact that the exchange
coupling is rather insensitive to the detailed orbital structure and 
chemical surrounding in transition metal oxides. 

\begin{figure}
{\epsfig{file=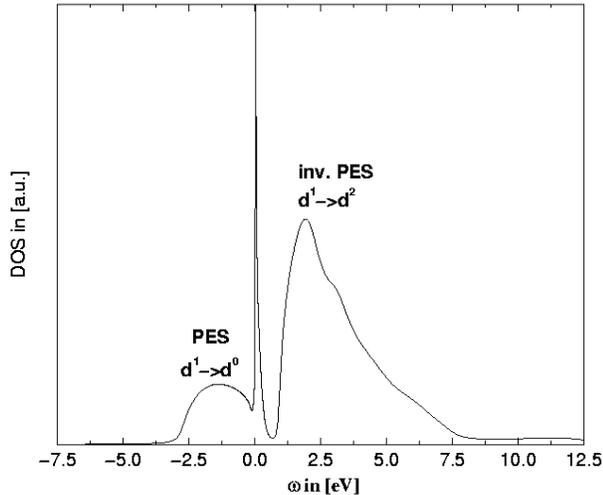,width=8cm}}  
\vspace{0.5cm}
\caption{The $t_{2g}$ density of states for LaTiO$_3$ for the orthorhombic 
structure and $\bar{U}=4 eV$ and $J=0.6 eV$  at a reciprocal temperature 
$\beta=100 \:eV^{-1}$.}    
\label{fig:3}
\end{figure}
The result of the LDA+DMFT calculation for the orthorhombic system is shown 
in figure \ref{fig:3}. The lowest energy difference between $d^1$ and $d^2$ 
is now reduced by $2 eV$ leading to a total shift of the upper Hubbard band 
compared with figure \ref{fig:2}.
The multiplet structure is not affected since it is proportional to the 
coupling constant $J$ only.  
For this set of parameters at a doping of $x=6\%$ the lower and the upper
Hubbard band is shifted towards the Fermi energy as a consequence of the 
smaller Coulomb interaction compared to the partially screened case.
The Kondo resonance is enlarged as a fact of different one-electron energy, 
which is the lowest for the orthorhombic set and for the full screened Coulomb 
energy $\bar{U}$. 

\begin{figure}
{\epsfig{file=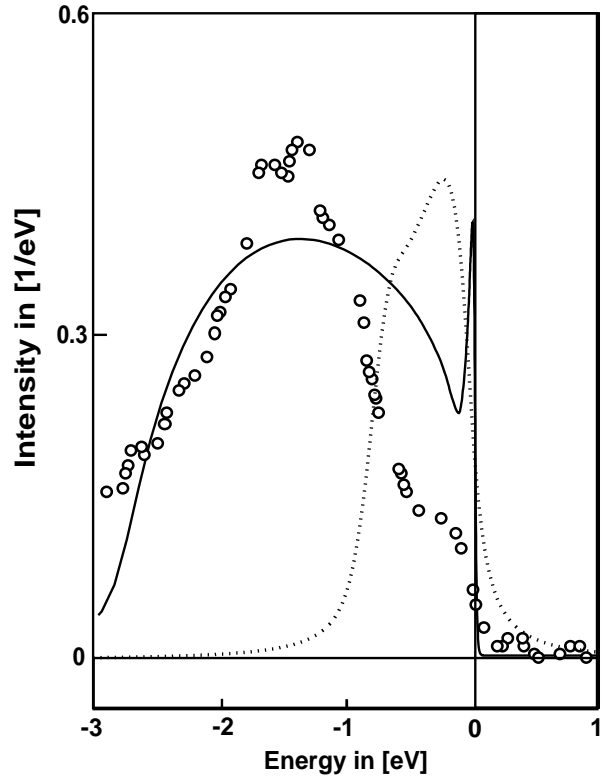,width=8cm}}
\vspace{0.5cm}  
\caption{The $t_{2g}$ density of states for LaTiO$_3$ for 
orthorhombic structure given by the LDA+DMFT calculation (full line) 
compared with LDA results (dotted curve) and data points from a photoemission 
experiment for $\rm La_{1-x}Sr_xTiO_3$ at a doping of $x=6\%$ 
(circles)\protect\cite{Fujimori}. 
The parameters of the calculation were 
$\bar{U}=4 eV$ and $J=0.6 eV$ and a reciprocal temperature 
$\beta=100\:eV^{-1}$. } 
\label{fig:4}
\end{figure}
A comparison with the experiment is done in figure \ref{fig:4} 
using the spectrum of the orthorombic system.
The bandwidth, maximum and center of mass of the lower Hubbard band are in fair
agreement with experiment. The many-body resonance at the Fermi level in the
spectrum in figure \ref{fig:2} manifests itself as an additional structure
at the Fermi level as a consequence of the convolution with the Fermi 
function. Note that such a structure -- although less pronounced -- can bee seen in the experimental data, too.
There are, of course, still differences in the exact distribution of the spectral
weight between experiment and our theory, which possibly originate from our
approximation of a purely local self-energy. Nevertheless, compared to the
pure LDA result, which gives a completely wrong account of the spectrum, the
LDA+DMFT appears to capture the essentials of the physics in this material.

\section{Summary}
\label{summary}
In this paper we have described one possible realization of a combination
of density-functional theory with the LDA approximation and the recently
developed dynamical mean field theory to obtain a first-principles calculation
scheme for strongly correlated electron systems. To solve the DMFT equations
we used the Non-Crossing approximation, which allowed us to include all local Coulomb interactions in our calculations, which 
is important to reproduce the full multiplet structure in the atomic 
limit. In order to test the method, we showed results for the doped Mott-Hubbard
insulator $\rm La_{1-x}Sr_xTiO_3$. The comparison of our results with a 
photoemission experiment shows, that the lower Hubbard band is of the correct
width and the center of mass of both experiment and theory is at the same 
position. Also, a shoulder seen at the Fermi energy in experiment is accounted
for by our results as the onset of a many-body resonance. In addition, the method appears to be very sensitive to the atomic
parameters and is clearly able to discriminate between different crystal structures and symmetries,
although the lattice enters the DMFT only in an averaged manner.
In total the results are very encouraging for the perspective of future 
applications.\\

\section*{Acknowledgements}
This work was partially supported by the DFG grant PR 289/5/1\&2.


%
%


\begin{references}
\bibitem{LSDA} P.\ Hohenberg, W.\ Kohn, Phys. Rev. {\bf 136,} B864 (1964);
W.\ Kohn, L.\ J.\ Sham, Phys. Rev. {\bf 140,} A1133 (1965).
\bibitem{LDA+U} V.\ I.\ Anisimov, J.\ Zaanen, O.\ K.\ Andersen, Phys. Rev. B {44,} 943 (1991).
\bibitem{LDA+Usuccess} V.\ I.\ Anisimov, F.\ Aryasetiawan, A.\ I.\ Lichtenstein, J. Phys.: Condens. Matter {\bf 9,} 767 (1997).
\bibitem{AniKot}V.\ I.\ Anisimov, A.\ I.\ Poteryaev, M.\ A.\ Korotin, A.\ O.\
Anokin, G.\ Kotliar, J. Phys.: Cond. Matter {\bf 9,} 7359 (1997). 
\bibitem{DMFT} Th.\ Pruschke, M.\ Jarrell and J.\ K.\ Freericks, Adv. in Phys. \textbf{44,} 187 (1995);
A.\ Georges, G.\ Kotliar W.\ Krauth, M.\ J.\ Rozenberg, Rev. Mod. Phys. \textbf{68,} 13 (1996).
\bibitem{QMCDMFT} M.\ Jarrell, Phys. Rev. Lett. {\bf 69,} 168 (1992);
M.\ Rozenberg, X.\ Y.\ Zhang, G.\ Kotliar, Phys. Rev. Lett. {\bf 69,} 
1236 (1992);
A.\ Georges, W.\ Krauth, Phys. Rev. Lett. {\bf 69,} 1240 (1992).
\bibitem{EDDMFT} M.\ Caffarel, W.\ Krauth, Phys. Rev. Lett. {\bf 72,} 1545 (1994);
M.\ Rozenberg, G.\ Moeller, G.\ Kotliar, Mod. Phys. Lett. B {\bf 8,} 535 (1994);
Q.\ Si, M.\ Rozenberg, G.\ Kotliar, A.\ E.\ Ruckenstein, Phys. Rev. Lett. {\bf 72,} 2761 (1994).
\bibitem{IPTDMFT}
A.\ Georges, G.\ Kotliar, Phys. Rev. B {\bf 45,} 6479 (1992).
\bibitem{NCADMFT} 
H.\ Keiter, J.\ C.\ Kimbal, Phys. Rev. Lett. {\bf 25,} 672 (1970);
N.\ E.\ Bickers, D.\ L.\ Cox, J.\ W.\ Wilkins, Phys. Rev. B {\bf 36,} 2036 (1987).
\bibitem{MOHM} 
K.\ Held, D.\ Vollhard, Euro. Phys. J. B {\bf 5,} 473 (1998). 
\bibitem{LMTO}O.\ K.\ Andersen, Phys. Rev. B {\bf 12,} 3060 (1975);
O.\ Gunnarsson, O.\ Jepsen, O.\ K.\ Andersen, Phys. Rev. B {\bf 27,}
7144 (1983).
\bibitem{Parameters} O.\ Gunnarson, O.\ K.\ Andersen, O.\ Jepsen, J.\ Zaanen, 
Phys. Rev. B {\bf 39,} 1708 (1989).
\bibitem{solovyev} I.Solovyev, N.Hamada, K.Terakura, Phys. Rev. B {\bf 53,}
7158 (1996).
\bibitem{pickett} W.E.Pickett, S.C.Erwin, E.C.Ethridge,
Phys. Rev. B {\bf 58,}1201 (1998).
\bibitem{Fujimori} A.\ Fujimori, et al., Phys. Rev B {\bf 46,} 9841 (1992).
(Actually, in this article, the chemical formula of the sample was 
$\rm LaTiO_{3.03}$, but the excess of oxygen produces $6\%$ holes, 
which is equivalent to doping with $6\%$ Sr.) 
\bibitem{Kondo} A.\ C.\ Hewson: The Kondo Problem to Heavy Fermions, 
Cambridge University Press, Cambridge (1993).
\end{references}
\end{document}